\documentclass[manuscript]{acmart}

\usepackage{booktabs}
\usepackage{multirow}

\AtBeginDocument{%
  \providecommand\BibTeX{{%
    \normalfont B\kern-0.5em{\scshape i\kern-0.25em b}\kern-0.8em\TeX}}}

\setcopyright{rightsretained}

\begin{document}

\copyrightyear{2020}
\acmYear{2020}
\acmConference[RecSys '20]{Fourteenth ACM Conference on Recommender Systems}{September 21--26, 2020}{Virtual Event, Brazil}
\acmBooktitle{Fourteenth ACM Conference on Recommender Systems (RecSys '20), September 21--26, 2020, Virtual Event, Brazil}
\acmDOI{10.1145/3383313.3411467}
\acmISBN{978-1-4503-7583-2/20/09}

\title[Auto-Surprise: An AutoRecSys Library]{Auto-Surprise: An Automated Recommender-System (AutoRecSys) Library with Tree of Parzens Estimator (TPE) Optimization}

\author{Rohan Anand}
\email{anandr@tcd.ie}
\author{Joeran Beel}
\email{beelj@scss.tcd.ie}
\affiliation{%
  \institution{Trinity College Dublin}
  \city{Dublin}
  \country{Ireland}
}

\renewcommand{\shortauthors}{Anand and Beel}

\begin{abstract}
We introduce Auto-Surprise\footnote{Source code available at \url{https://github.com/BeelGroup/Auto-Surprise}. For full documentation, see \url{https://auto-surprise.readthedocs.io/en/stable/}}, an automated recommender system library. Auto-Surprise is an extension of the Surprise recommender system library and eases the algorithm selection and configuration process. Compared to an out-of-the-box Surprise library, without hyper parameter optimization, AutoSurprise performs better, when evaluated with MovieLens, Book Crossing and Jester datasets. It may also result in the selection of an algorithm with significantly lower runtime. Compared to Surprise’s grid search, Auto-Surprise performs equally well or slightly better in terms of RMSE, and is notably faster in finding the optimum hyperparameters. 
\end{abstract}



\maketitle

\section{Introduction}
Recommender-system development has always been a challenge. Particularly, identifying the best recommendation algorithm and parameters for a given scenario is difficult. `Intuition', even of experienced data scientists, is often not good enough to identify the ideal algorithm and parameters \cite{gomez2015netflix}. Minor variations in implementations and parameters may lead to significantly different performances in different scenarios \cite{beel2016towards}.

The machine learning community faces similar challenges and tackled these quite successfully with so-called Automated Machine Learning (AutoML). AutoML eases the configuration of machine learning pipelines, particularly the algorithm selection and configuration process. AutoML applies hyperparameter optimization techniques beyond standard grid or random search, not only to hyperparameters but also to algorithm selection \cite{hutter2019automated}. Typical AutoML methods include Bayesian optimization \cite{snoek2012practical}, Sequential model-based optimization \cite{hutter2011sequential} or hierarchical planning \cite{mohr2018ml}. Sometimes, metalearning is used to ‘warm-start’ the process, i.e. to predict a set of algorithms and parameters that are promising for a given task \cite{hutter2019automated}. AutoML is easily accessible for machine-learning engineers through AutoML software libraries including H2O \cite{candel2016deep}, TPOT \cite{olson2019tpot}, AutoWEKA \cite{thornton2013auto}, AutoSklearn \cite{feurer2015efficient}, AutoKeras \cite{jin2019auto}, and MLPlan \cite{mohr2018ml}. 

The recommender-system community has fallen behind the advances in the (automated) machine-learning community. While there are many recommender-system libraries such as Mahout \cite{lyubimov2016apache}, LibRec \cite{guo2015librec}, Surprise \cite{hug2017surprise}, CaseRec \cite{da2018case}, and Lenskit \cite{ekstrand2011rethinking}, there is – to the best of our knowledge – only one Automated Recommender System library, namely Librec-Auto \cite{mansoury2019algorithm}. Librec-Auto extends the LibRec recommender-system library with some automated algorithm selection and configuration functionality, though this functionality is limited. LibRec-Auto iterates over parameter spaces in one scripted experiment, whereas the user still must define the parameter spaces and write the script. This system is useful for experienced data scientists who wish to experiment with different configurations and analyze the models. However, LibRec-Auto is not as advanced as the typical AutoML tools, and a user with no prior experience may have difficulties in setting up such a solution.

We introduce Auto-Surprise\footnotemark[1], the first automated recommender system library (AutoRecSys) with fully automated algorithm selection and configuration, comparable to state of-the-art AutoML libraries.

\section{Auto-Surprise}

Auto-Surprise is built as a wrapper around the Python Surprise \cite{hug2017surprise} library. Auto-Surprise uses a sequential model-based optimization approach for the algorithm selection and configuration, is open-source and brings the advances of AutoML to the recommender-system community. Auto-Surprise offers all 11 algorithms (see Table \ref{tab:results}) that Surprise has implemented. To use Auto-Surprise, a user needs to import the auto-surprise package and pass data to the trainer method. Auto-Surprise then automatically identifies the best performing algorithm and hyperparameters out of the 11 algorithms. As such, almost no prior knowledge is needed. 

\begin{figure}[h]
  \centering
  \includegraphics[width=.7\linewidth]{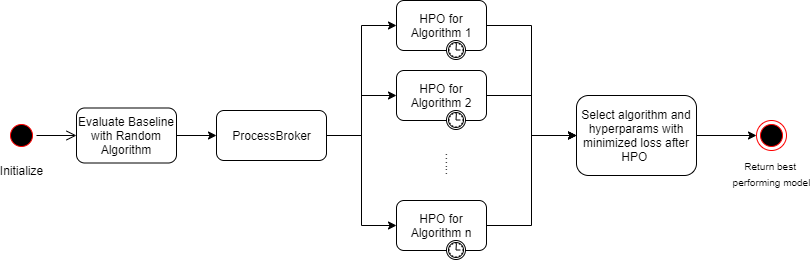}
  \caption{Simple overview of the working of Auto-Surprise}
  \Description{A simple overview of the working of Auto-Surprise}
\end{figure}

The overall optimization strategy of Auto-Surprise is similar to AutoWEKA \cite{thornton2013auto}. Auto-Surprise first evaluate a baseline score for the given dataset using random predictor. This sets the minimum loss that each algorithm must achieve. Each algorithm is then optimized in parallel until a user defined time limit or a maximum evaluations limit is reached. If any of the algorithms perform worse than the baseline after a number of evaluations, it is not optimized any further. Once this process is completed, the best performing algorithm with optimized hyperparameters is returned along with a dictionary of the performance of all the algorithm.

For all algorithms (except for those which do not require any hyperparameters), we defined a hyperparameter space which is used to identify optimal hyperparameters in the given range. Auto-Surprise can use three hyperparameter optimization methods as implemented by Hyperopt \cite{bergstra2013hyperopt} - Tree of Parzens Estimator (TPE) \cite{bergstra2011algorithms}, Adaptive TPE (ATPE) \cite{hypermaxAtpe} and Random Search. The user sets a target metric such as RMSE or MAE which is to be minimized. All of this is done in just one line of code.
 
\section{Evaluation}

We compared Auto-Surprise against all eleven algorithms in Surprise with a) the algorithms' default parameters and b) the algorithms' being optimized with Grid search as implemented by Surprise with concurrency enabled. We used the Movielens 100k dataset \cite{harper2015movielens}, the de-facto gold-standard dataset in the recommender system community \cite{beel2019data}. Jester-2 \cite{goldberg2001eigentaste} and Book Crossing \cite{ziegler2005improving} datasets were also used, though only using a 100k sample of them to reduce resource requirements. Separate configurations of Auto-Surprise were evaluated using Adaptive TPE and TPE as the hyperparameter optimization algorithm. The target metric to minimize was RMSE and a maximum evaluation time of 2 hours was set for Auto-Surprise.

\begin{table*}
\caption{Comparison of Auto-Surprise with other Surprise algorithms and Grid Search. Results in bold is for the overall best performing algorithm in its default configuration}
\label{tab:results}
\centering
\begin{tabular}{lccccccccc}
\hline
\multicolumn{1}{c}{\multirow{2}{*}{\textbf{Algorithm}}} & \multicolumn{3}{c}{\textbf{MovieLens 100k}}                      & \multicolumn{3}{c}{\textbf{Jester 2}}                          & \multicolumn{3}{c}{\textbf{Book Crossing}}                      \\ \cline{2-10} 
\multicolumn{1}{c}{}                                    & \textbf{RMSE}   & \textbf{MAE}    & \textbf{Time}                & \textbf{RMSE}  & \textbf{MAE}   & \textbf{Time}                & \textbf{RMSE}   & \textbf{MAE}   & \textbf{Time}                \\ \hline
Normal Predictor                                        & 1.5195          & 1.2200          & 00:00:01                     & 7.277          & 5.886          & 00:00:01                     & 4.8960          & 3.866          & 00:00:01                     \\
SVD                                                     & 0.9364          & 0.7385          & 00:00:23                     & 4.905          & 3.97           & 00:00:13                     & 3.5586          & 3.013          & 00:00:11                     \\
SVD++                                                   & 0.9196          & 0.7216          & 00:14:23                     & 5.102          & 4.055          & 00:00:29                     & 3.5842          & 2.991          & 00:01:48                     \\
NMF                                                     & 0.9651          & 0.7592          & 00:00:25                     & --             & --             & --                           & --              & --             & --                           \\
Slope One                                               & 0.9450          & 0.7425          & 00:00:15                     & 5.189          & 3.945          & 00:00:02                     & --              & --             & --                           \\
KNN Basic                                               & 0.9791          & 0.7738          & 00:00:18                     & 5.078          & 4.034          & 00:02:14                     & 3.9108          & 3.562          & 00:00:38                     \\
KNN with Means                                          & 0.9510          & 0.7490          & 00:00:19                     & 5.124          & 3.955          & 00:02:16                     & 3.8574          & 3.301          & 00:00:35                     \\
KNN with   Z-score                                      & 0.9517          & 0.7470          & 00:00:21                     & 5.219          & 3.955          & 00:02:20                     & 3.8526          & 3.292          & 00:00:37                     \\
\textbf{KNN Baseline}                                   & \textbf{0.9299} & \textbf{0.7329} & \textbf{00:00:22}            & \textbf{4.898} & \textbf{3.896} & \textbf{00:02:14}            & \textbf{3.6181} & \textbf{3.101} & \textbf{00:00:36}            \\
Co-clustering                                           & 0.9678          & 0.7581          & 00:00:08                     & 5.153          & 3.917          & 00:00:12                     & 4.0168          & 3.409          & 00:00:19                     \\
Baseline Only                                           & 0.9433          & 0.7479          & 00:00:01                     & 4.849          & 3.934          & 00:00:01                     & 3.5760          & 3.095          & 00:00:02                     \\ \hline
GridSearch                                              & 0.9139          & 0.7167          & \multicolumn{1}{l}{27:02:48} & 4.7409         & 3.8147         & \multicolumn{1}{l}{80:52:35} & 3.5467          & 2.9554         & \multicolumn{1}{l}{48:29:46} \\
Auto-Surprise (TPE)                                     & 0.9136          & 0.7280          & 02:00:01                     & 4.6489         & 3.6837         & 02:00:10                     & 3.5221          & 2.8871         & 02:00:58                     \\
Auto-Surprise (ATPE)                                    & 0.9116          & 0.7244          & 02:00:02                     & 4.6555         & 3.6906         & 02:00:01                     & 3.5190          & 2.8739         & 02:00:06                     \\ \hline
\end{tabular}
\end{table*}

The best default algorithm in Surprise for the MovieLens dataset was SVD++ with an RMSE of 0.9196. Auto-Surprise was able to perform best with adaptive TPE with an RMSE of 0.9116. This is a small - but statistically significant difference (2 tailed p value < 0.05) in RMSE of 0.86\%. We see a similar result for the Book Crossing dataset which is optimized from an RMSE of 3.5586 with SVD to 3.5190 with Auto-Surprise, a 1.11\% difference. However, we see a more pronounced difference for the Jester dataset from Baseline Only algorithm with an RMSE of 4.8490 and Auto-Surprise with an RMSE of 4.6489, a difference of 4.12\%. It is also important to note that the best performing algorithm by default may not be the algorithm selected after optimization. For Movielens, Auto-Surprise found that NMF performed best after optimization even though the best performing default algorithm was SVD++. Similarly with Jester, KNN Baseline was selected and for Book Crossing, SVD was selected. While GridSearch can result in decent results, the time taken is far longer than Auto-Surprise.

\section{Conclusion}
We found that when compared to the default configuration of Surprise algorithms, Auto-Surprise performs anywhere from 0.8\% to 4\% better in terms of RMSE in our tests. Though the actual run time of the combined default Surprise algorithms is still lower than Auto-Surprise, it still widely outperforms Gridsearch in that respect. It is also worth noting that the selected algorithm may have a much lower runtime compared to the default algorithm as shown for the Movielens dataset where the selected algorithm NMF only has a runtime of 25 seconds compared to the ~15 minutes runtime for SVD++, the best performing default algorithm. And, of course, Auto-Surprise eases the entire process of algorithm selection and hyperparameter optimization by automating it in a single line of code.


\bibliographystyle{ACM-Reference-Format}
\bibliography{ref}


\begin{thebibliography}{24}


\ifx \showCODEN    \undefined \def \showCODEN     #1{\unskip}     \fi
\ifx \showDOI      \undefined \def \showDOI       #1{#1}\fi
\ifx \showISBNx    \undefined \def \showISBNx     #1{\unskip}     \fi
\ifx \showISBNxiii \undefined \def \showISBNxiii  #1{\unskip}     \fi
\ifx \showISSN     \undefined \def \showISSN      #1{\unskip}     \fi
\ifx \showLCCN     \undefined \def \showLCCN      #1{\unskip}     \fi
\ifx \shownote     \undefined \def \shownote      #1{#1}          \fi
\ifx \showarticletitle \undefined \def \showarticletitle #1{#1}   \fi
\ifx \showURL      \undefined \def \showURL       {\relax}        \fi
\providecommand\bibfield[2]{#2}
\providecommand\bibinfo[2]{#2}
\providecommand\natexlab[1]{#1}
\providecommand\showeprint[2][]{arXiv:#2}

\bibitem[\protect\citeauthoryear{Beel, Breitinger, Langer, Lommatzsch, and
  Gipp}{Beel et~al\mbox{.}}{2016}]%
        {beel2016towards}
\bibfield{author}{\bibinfo{person}{Joeran Beel}, \bibinfo{person}{Corinna
  Breitinger}, \bibinfo{person}{Stefan Langer}, \bibinfo{person}{Andreas
  Lommatzsch}, {and} \bibinfo{person}{Bela Gipp}.}
  \bibinfo{year}{2016}\natexlab{}.
\newblock \showarticletitle{Towards reproducibility in recommender-systems
  research}.
\newblock \bibinfo{journal}{\emph{User modeling and user-adapted interaction}}
  \bibinfo{volume}{26}, \bibinfo{number}{1} (\bibinfo{year}{2016}),
  \bibinfo{pages}{69--101}.
\newblock


\bibitem[\protect\citeauthoryear{Beel and Brunel}{Beel and Brunel}{2019}]%
        {beel2019data}
\bibfield{author}{\bibinfo{person}{Joeran Beel} {and} \bibinfo{person}{Victor
  Brunel}.} \bibinfo{year}{2019}\natexlab{}.
\newblock \showarticletitle{Data Pruning in Recommender Systems Research:
  Best-Practice or Malpractice}.
\newblock \bibinfo{journal}{\emph{ACM RecSys}} (\bibinfo{year}{2019}).
\newblock


\bibitem[\protect\citeauthoryear{Bergstra, Yamins, and Cox}{Bergstra
  et~al\mbox{.}}{2013}]%
        {bergstra2013hyperopt}
\bibfield{author}{\bibinfo{person}{James Bergstra}, \bibinfo{person}{Dan
  Yamins}, {and} \bibinfo{person}{David~D Cox}.}
  \bibinfo{year}{2013}\natexlab{}.
\newblock \showarticletitle{Hyperopt: A python library for optimizing the
  hyperparameters of machine learning algorithms}. In
  \bibinfo{booktitle}{\emph{Proceedings of the 12th Python in science
  conference}}. Citeseer, \bibinfo{pages}{13--20}.
\newblock


\bibitem[\protect\citeauthoryear{Bergstra, Bardenet, Bengio, and
  K{\'e}gl}{Bergstra et~al\mbox{.}}{2011}]%
        {bergstra2011algorithms}
\bibfield{author}{\bibinfo{person}{James~S Bergstra}, \bibinfo{person}{R{\'e}mi
  Bardenet}, \bibinfo{person}{Yoshua Bengio}, {and} \bibinfo{person}{Bal{\'a}zs
  K{\'e}gl}.} \bibinfo{year}{2011}\natexlab{}.
\newblock \showarticletitle{Algorithms for hyper-parameter optimization}. In
  \bibinfo{booktitle}{\emph{Advances in neural information processing
  systems}}. \bibinfo{pages}{2546--2554}.
\newblock


\bibitem[\protect\citeauthoryear{Candel, Parmar, LeDell, and Arora}{Candel
  et~al\mbox{.}}{2016}]%
        {candel2016deep}
\bibfield{author}{\bibinfo{person}{Arno Candel}, \bibinfo{person}{Viraj
  Parmar}, \bibinfo{person}{Erin LeDell}, {and} \bibinfo{person}{Anisha
  Arora}.} \bibinfo{year}{2016}\natexlab{}.
\newblock \showarticletitle{Deep learning with H2O}.
\newblock \bibinfo{journal}{\emph{H2O. ai Inc}} (\bibinfo{year}{2016}).
\newblock


\bibitem[\protect\citeauthoryear{da~Costa, Fressato, Neto, Manzato, and
  Campello}{da~Costa et~al\mbox{.}}{2018}]%
        {da2018case}
\bibfield{author}{\bibinfo{person}{Arthur da Costa}, \bibinfo{person}{Eduardo
  Fressato}, \bibinfo{person}{Fernando Neto}, \bibinfo{person}{Marcelo
  Manzato}, {and} \bibinfo{person}{Ricardo Campello}.}
  \bibinfo{year}{2018}\natexlab{}.
\newblock \showarticletitle{Case recommender: a flexible and extensible python
  framework for recommender systems}. In \bibinfo{booktitle}{\emph{Proceedings
  of the 12th ACM Conference on Recommender Systems}}.
  \bibinfo{pages}{494--495}.
\newblock


\bibitem[\protect\citeauthoryear{Ekstrand, Ludwig, Konstan, and Riedl}{Ekstrand
  et~al\mbox{.}}{2011}]%
        {ekstrand2011rethinking}
\bibfield{author}{\bibinfo{person}{Michael~D Ekstrand},
  \bibinfo{person}{Michael Ludwig}, \bibinfo{person}{Joseph~A Konstan}, {and}
  \bibinfo{person}{John~T Riedl}.} \bibinfo{year}{2011}\natexlab{}.
\newblock \showarticletitle{Rethinking the recommender research ecosystem:
  reproducibility, openness, and LensKit}. In
  \bibinfo{booktitle}{\emph{Proceedings of the fifth ACM conference on
  Recommender systems}}. \bibinfo{pages}{133--140}.
\newblock


\bibitem[\protect\citeauthoryear{Electricbrain}{Electricbrain}{2019}]%
        {hypermaxAtpe}
\bibfield{author}{\bibinfo{person}{Electricbrain}.}
  \bibinfo{year}{2019}\natexlab{}.
\newblock \bibinfo{title}{Hypermax}.
\newblock
  \bibinfo{howpublished}{\url{https://github.com/electricbrainio/hypermax}}.
\newblock


\bibitem[\protect\citeauthoryear{Feurer, Klein, Eggensperger, Springenberg,
  Blum, and Hutter}{Feurer et~al\mbox{.}}{2015}]%
        {feurer2015efficient}
\bibfield{author}{\bibinfo{person}{Matthias Feurer}, \bibinfo{person}{Aaron
  Klein}, \bibinfo{person}{Katharina Eggensperger}, \bibinfo{person}{Jost
  Springenberg}, \bibinfo{person}{Manuel Blum}, {and} \bibinfo{person}{Frank
  Hutter}.} \bibinfo{year}{2015}\natexlab{}.
\newblock \showarticletitle{Efficient and robust automated machine learning}.
  In \bibinfo{booktitle}{\emph{Advances in neural information processing
  systems}}. \bibinfo{pages}{2962--2970}.
\newblock


\bibitem[\protect\citeauthoryear{Goldberg, Roeder, Gupta, and Perkins}{Goldberg
  et~al\mbox{.}}{2001}]%
        {goldberg2001eigentaste}
\bibfield{author}{\bibinfo{person}{Ken Goldberg}, \bibinfo{person}{Theresa
  Roeder}, \bibinfo{person}{Dhruv Gupta}, {and} \bibinfo{person}{Chris
  Perkins}.} \bibinfo{year}{2001}\natexlab{}.
\newblock \showarticletitle{Eigentaste: A constant time collaborative filtering
  algorithm}.
\newblock \bibinfo{journal}{\emph{information retrieval}} \bibinfo{volume}{4},
  \bibinfo{number}{2} (\bibinfo{year}{2001}), \bibinfo{pages}{133--151}.
\newblock


\bibitem[\protect\citeauthoryear{Gomez-Uribe and Hunt}{Gomez-Uribe and
  Hunt}{2015}]%
        {gomez2015netflix}
\bibfield{author}{\bibinfo{person}{Carlos~A Gomez-Uribe} {and}
  \bibinfo{person}{Neil Hunt}.} \bibinfo{year}{2015}\natexlab{}.
\newblock \showarticletitle{The netflix recommender system: Algorithms,
  business value, and innovation}.
\newblock \bibinfo{journal}{\emph{ACM Transactions on Management Information
  Systems (TMIS)}} \bibinfo{volume}{6}, \bibinfo{number}{4}
  (\bibinfo{year}{2015}), \bibinfo{pages}{1--19}.
\newblock


\bibitem[\protect\citeauthoryear{Guo, Zhang, Sun, and Yorke-Smith}{Guo
  et~al\mbox{.}}{2015}]%
        {guo2015librec}
\bibfield{author}{\bibinfo{person}{Guibing Guo}, \bibinfo{person}{Jie Zhang},
  \bibinfo{person}{Zhu Sun}, {and} \bibinfo{person}{Neil Yorke-Smith}.}
  \bibinfo{year}{2015}\natexlab{}.
\newblock \showarticletitle{LibRec: A Java Library for Recommender Systems.}.
  In \bibinfo{booktitle}{\emph{UMAP Workshops}}, Vol.~\bibinfo{volume}{4}.
\newblock


\bibitem[\protect\citeauthoryear{Harper and Konstan}{Harper and
  Konstan}{2015}]%
        {harper2015movielens}
\bibfield{author}{\bibinfo{person}{F~Maxwell Harper} {and}
  \bibinfo{person}{Joseph~A Konstan}.} \bibinfo{year}{2015}\natexlab{}.
\newblock \showarticletitle{The movielens datasets: History and context}.
\newblock \bibinfo{journal}{\emph{Acm transactions on interactive intelligent
  systems (tiis)}} \bibinfo{volume}{5}, \bibinfo{number}{4}
  (\bibinfo{year}{2015}), \bibinfo{pages}{1--19}.
\newblock


\bibitem[\protect\citeauthoryear{Hug}{Hug}{2017}]%
        {hug2017surprise}
\bibfield{author}{\bibinfo{person}{Nicolas Hug}.}
  \bibinfo{year}{2017}\natexlab{}.
\newblock \showarticletitle{Surprise, a Python library for recommender
  systems}.
\newblock \bibinfo{journal}{\emph{URL: http://surpriselib. com}}
  (\bibinfo{year}{2017}).
\newblock


\bibitem[\protect\citeauthoryear{Hutter, Hoos, and Leyton-Brown}{Hutter
  et~al\mbox{.}}{2011}]%
        {hutter2011sequential}
\bibfield{author}{\bibinfo{person}{Frank Hutter}, \bibinfo{person}{Holger~H
  Hoos}, {and} \bibinfo{person}{Kevin Leyton-Brown}.}
  \bibinfo{year}{2011}\natexlab{}.
\newblock \showarticletitle{Sequential model-based optimization for general
  algorithm configuration}. In \bibinfo{booktitle}{\emph{International
  conference on learning and intelligent optimization}}. Springer,
  \bibinfo{pages}{507--523}.
\newblock


\bibitem[\protect\citeauthoryear{Hutter, Kotthoff, and Vanschoren}{Hutter
  et~al\mbox{.}}{2019}]%
        {hutter2019automated}
\bibfield{author}{\bibinfo{person}{Frank Hutter}, \bibinfo{person}{Lars
  Kotthoff}, {and} \bibinfo{person}{Joaquin Vanschoren}.}
  \bibinfo{year}{2019}\natexlab{}.
\newblock \bibinfo{booktitle}{\emph{Automated Machine Learning}}.
\newblock \bibinfo{publisher}{Springer}.
\newblock


\bibitem[\protect\citeauthoryear{Jin, Song, and Hu}{Jin et~al\mbox{.}}{2019}]%
        {jin2019auto}
\bibfield{author}{\bibinfo{person}{Haifeng Jin}, \bibinfo{person}{Qingquan
  Song}, {and} \bibinfo{person}{Xia Hu}.} \bibinfo{year}{2019}\natexlab{}.
\newblock \showarticletitle{Auto-keras: An efficient neural architecture search
  system}. In \bibinfo{booktitle}{\emph{Proceedings of the 25th ACM SIGKDD
  International Conference on Knowledge Discovery \& Data Mining}}.
  \bibinfo{pages}{1946--1956}.
\newblock


\bibitem[\protect\citeauthoryear{Lyubimov and Palumbo}{Lyubimov and
  Palumbo}{2016}]%
        {lyubimov2016apache}
\bibfield{author}{\bibinfo{person}{Dmitriy Lyubimov} {and}
  \bibinfo{person}{Andrew Palumbo}.} \bibinfo{year}{2016}\natexlab{}.
\newblock \bibinfo{booktitle}{\emph{Apache Mahout: Beyond MapReduce}}.
\newblock \bibinfo{publisher}{CreateSpace Independent Publishing Platform}.
\newblock


\bibitem[\protect\citeauthoryear{Mansoury and Burke}{Mansoury and
  Burke}{2019}]%
        {mansoury2019algorithm}
\bibfield{author}{\bibinfo{person}{Masoud Mansoury} {and}
  \bibinfo{person}{Robin Burke}.} \bibinfo{year}{2019}\natexlab{}.
\newblock \showarticletitle{Algorithm Selection with Librec-auto.}. In
  \bibinfo{booktitle}{\emph{AMIR@ ECIR}}. \bibinfo{pages}{11--17}.
\newblock


\bibitem[\protect\citeauthoryear{Mohr, Wever, and H{\"u}llermeier}{Mohr
  et~al\mbox{.}}{2018}]%
        {mohr2018ml}
\bibfield{author}{\bibinfo{person}{Felix Mohr}, \bibinfo{person}{Marcel Wever},
  {and} \bibinfo{person}{Eyke H{\"u}llermeier}.}
  \bibinfo{year}{2018}\natexlab{}.
\newblock \showarticletitle{ML-Plan: Automated machine learning via
  hierarchical planning}.
\newblock \bibinfo{journal}{\emph{Machine Learning}} \bibinfo{volume}{107},
  \bibinfo{number}{8-10} (\bibinfo{year}{2018}), \bibinfo{pages}{1495--1515}.
\newblock


\bibitem[\protect\citeauthoryear{Olson and Moore}{Olson and Moore}{2019}]%
        {olson2019tpot}
\bibfield{author}{\bibinfo{person}{Randal~S Olson} {and}
  \bibinfo{person}{Jason~H Moore}.} \bibinfo{year}{2019}\natexlab{}.
\newblock \showarticletitle{TPOT: A tree-based pipeline optimization tool for
  automating machine learning}.
\newblock In \bibinfo{booktitle}{\emph{Automated Machine Learning}}.
  \bibinfo{publisher}{Springer}, \bibinfo{pages}{151--160}.
\newblock


\bibitem[\protect\citeauthoryear{Snoek, Larochelle, and Adams}{Snoek
  et~al\mbox{.}}{2012}]%
        {snoek2012practical}
\bibfield{author}{\bibinfo{person}{Jasper Snoek}, \bibinfo{person}{Hugo
  Larochelle}, {and} \bibinfo{person}{Ryan~P Adams}.}
  \bibinfo{year}{2012}\natexlab{}.
\newblock \showarticletitle{Practical bayesian optimization of machine learning
  algorithms}. In \bibinfo{booktitle}{\emph{Advances in neural information
  processing systems}}. \bibinfo{pages}{2951--2959}.
\newblock


\bibitem[\protect\citeauthoryear{Thornton, Hutter, Hoos, and
  Leyton-Brown}{Thornton et~al\mbox{.}}{2013}]%
        {thornton2013auto}
\bibfield{author}{\bibinfo{person}{Chris Thornton}, \bibinfo{person}{Frank
  Hutter}, \bibinfo{person}{Holger~H Hoos}, {and} \bibinfo{person}{Kevin
  Leyton-Brown}.} \bibinfo{year}{2013}\natexlab{}.
\newblock \showarticletitle{Auto-WEKA: Combined selection and hyperparameter
  optimization of classification algorithms}. In
  \bibinfo{booktitle}{\emph{Proceedings of the 19th ACM SIGKDD international
  conference on Knowledge discovery and data mining}}.
  \bibinfo{pages}{847--855}.
\newblock


\bibitem[\protect\citeauthoryear{Ziegler, McNee, Konstan, and Lausen}{Ziegler
  et~al\mbox{.}}{2005}]%
        {ziegler2005improving}
\bibfield{author}{\bibinfo{person}{Cai-Nicolas Ziegler},
  \bibinfo{person}{Sean~M McNee}, \bibinfo{person}{Joseph~A Konstan}, {and}
  \bibinfo{person}{Georg Lausen}.} \bibinfo{year}{2005}\natexlab{}.
\newblock \showarticletitle{Improving recommendation lists through topic
  diversification}. In \bibinfo{booktitle}{\emph{Proceedings of the 14th
  international conference on World Wide Web}}. \bibinfo{pages}{22--32}.
\newblock


\end{thebibliography}

\end{document}